\documentstyle [aps,multicol,psfig]{revtex}
\renewcommand{\narrowtext}{\begin{multicols}{2} \global\columnwidth20.5pc}
\renewcommand{\widetext}{\end{multicols} \global\columnwidth42.5pc}

\begin{document}

\newcommand{\newc}{\newcommand}

\newc{\be}{\begin{equation}}
\newc{\ee}{\end{equation}}
\newc{\ba}{\begin{eqnarray}}
\newc{\ea}{\end{eqnarray}}
\newc{\bea}{\begin{eqnarray*}}
\newc{\eea}{\end{eqnarray*}}
\newc{\D}{\partial}
\newc{\ie}{{\it i.e.} }
\newc{\eg}{{\it e.g.} }
\newc{\etc}{{\it etc.} }
\newc{\etal}{{\it et al.}}

\newc{\ra}{\rightarrow}
\newc{\lra}{\leftrightarrow}
\newc{\no}{Nielsen-Olesen }
\newc{\lsim}{\buildrel{<}\over{\sim}}
\newc{\gsim}{\buildrel{>}\over{\sim}}

\title{Dynamics of Nontopological Solitons - Q Balls
\footnote{The animated simulations may be found on-line at the 
address http://leandros.chem.demokritos.gr/qballs/index.html}} 
\author{Minos Axenides$^a$, Stavros Komineas$^b$, Leandros Perivolaropoulos$^a$,
Manolis Floratos$^a$}
\address{$^a$ Institute of Nuclear Physics, N.C.R.P.S. Demokritos,  153
10, Athens, Greece \\ $^b$ Physikalisches Institut, Universit\"at
Bayreuth, D-95440 Bayreuth, Germany.} 

\date{October 18, 1999}
\maketitle

\begin{abstract}

We use numerical simulations and semi-analytical methods to
investigate the stability and the interactions of nontopological
stationary qball solutions. In the context of a simple model we
 map the parameter sectors of
stability for a single qball and verify the result using numerical
simulations of time evolution. The system of two interacting
qballs is also studied in one and two space dimensions. We find
that the system generically performs breather type oscillations
with frequency equal to the difference of the internal qball
frequencies. This result is shown to be consistent with the form
of the qball interaction potential. Finally we perform simulations
of qball scattering and show that the right angle scattering
effect observed in topological soliton scattering in two dimensions,
persists also in
the case of qballs where no topologically conserved quantities are
present. For relativistic collision velocities the qball charge is
split into a forward and a right angle scattering component.  As
the collision  velocity increases, the forward component gets
amplified at the expense of the right angle component.

\end{abstract}

\narrowtext

\section{Introduction}

It is well known that realistic supersymmetric theories are
associated with a number of scalar fields with various charges.
These theories and in particular the MSSM allow for baryonic and
leptonic nontopological solitons\cite{lp92} known as {\it
qballs}\cite{c85}. They are composed  of squarks, sleptons and
Higgs scalars \cite{k97,k98} and they originate through the 
breakup of an Affleck-Dine condensate carrying a net baryon and/or 
lepton number\cite{ks98}. Such objects could have interesting 
cosmological consequences. For example they could be responsible 
for both the net baryon number of the universe and its dark matter 
component\cite{k98}. 

The basic properties of qballs have been studied extensively in
the literature using mainly analytical and semi-analytical
methods\cite{lp92}. These methods are particularly useful in
understanding basic properties of qballs like existence, small
vibrations, stability in certain parameter limits (thick\cite{k97}
or thin wall\cite{c85} approximation) but they can not be very
illuminating in understanding more complicated issues like
scattering, interactions or stability for arbitrary parameters.
The goal of this paper is to use numerical simulations of qball
evolution in one and two spatial dimensions, along with
semi-analytical methods in order to study the stability under
small fluctuations, the interactions and the scattering of qballs.

The structure of the paper is the following: In the next section
we introduce a simple model and show the existence of stable qball
solutions in its context. By considering small fluctuations
superposed on these solutions for various parameters we construct
a map showing the stability sectors in parameter space. These
sectors are then verified by performing numerical simulations of
time evolution for the solutions considered. A virial theorem is
also derived and its validity is demonstrated numerically. In
section III we consider a system of two interacting qballs and
show analytically that the interaction potential is time dependent
and periodic. This type of interaction is verified by performing
numerical simulations of qball evolution which shows that the
system performs breather-type oscillations with the analytically
predicted frequency. In section IV we use boosted qball
configurations to perform scattering numerical experiments in 2+1
dimensions showing that for low collision velocities qballs tend
to scatter at right angles like their topological counterparts.
For relativistic collision velocities we find that the qball
charge splits into a forward and a right angle component. The
forward component gets amplified as the velocity increases.
Finally in section V we conclude, summarize and briefly discuss
future extensions of this work.

\section{Stability - Virial Theorem}

Consider a complex scalar field $\Phi$ in 1+1 dimensions whose
dynamics is determined by the Lagrangian
\begin{equation}
{\cal L} =  {1\over 2} \partial_\mu \Phi^* \partial^\mu \Phi  -
U(\Phi), \label{model}
\end{equation}
where
\be
U(\Phi)={1\over 2} m^2 |\Phi|^2 - {1\over 3} \alpha |\Phi|^3 +
{1\over 4} b |\Phi|^4.
\label{pot}
\ee
Using a rescaling
\ba
\Phi & \rightarrow & {m^2 \over \alpha} \Phi, \\
x & \rightarrow & {x\over m},
\ea
the Lagrangian (\ref{model}) gets simplified as follows
\begin{equation}
{\cal L} =  {1\over 2} \partial_\mu \Phi^* \partial^\mu \Phi  -
{1\over 2} |\Phi|^2 + {1\over 3} |\Phi|^3 - {1\over 4} B |\Phi|^4,
 \label{modelresc}
\end{equation}
where
\be
B\equiv {{b m^2}\over \alpha^2}. \ee
This leads to the field equation
\be
{\ddot \Phi} -\Phi'' + \Phi - |\Phi| \Phi + B |\Phi|^2 \Phi = 0.
\label{feq}\ee
Using the usual qball ansatz
\be
\Phi (x) = \sigma (x) e^{i \omega t} \label{ansatz} \ee with the
conserved Noether charge
\ba
Q & = & {1 \over 2 i}
  \int_{-\infty}^{+\infty}dx \;  (\Phi^*\, \D_t \Phi - \Phi\, \D_t \Phi^*)
 \nonumber \\
  & = & \omega \int_{-\infty}^{+\infty}dx \; \sigma (x)^2
\label{charge}
\ea
we obtain the field equation for $\sigma (x)$
\be
\sigma'' + (\omega^2 -1) \sigma + \sigma^2 - B \sigma^3 = 0.
\label{qeq} \ee The requirement of finite energy and the
asymptotics obtained from equation (\ref{qeq}) imply the
boundary conditions

\ba \sigma'(0)&=&0, \\ \sigma(\infty)&=& 0 \label{sigbou}. \ea

The combination of equations (\ref{qeq}) and (11-12) are identical
to the dynamical equations describing the motion of a virtual
particle that starts at rest at `time' $x=0$ from some position
$\sigma_0$ and moves to $\sigma=0$ at `time' $x\rightarrow \infty$
under the influence of the effective potential
\be
V_{eff}={1\over 2}(\omega^2-1) |\Phi|^2 + {1\over 3} |\Phi|^3 -
{1\over 4} B |\Phi|^4 \label{effpot} \ee.

By considering the form of the effective potential (\ref{effpot})  
it becomes clear that in order to have a solution with the 
boundary conditions (11-12) the following conditions must be 
satisfied 

\begin{itemize}
\item
Symmetry must be broken by the effective potential (\ref{effpot})
($m_{eff}^2 \equiv \omega^2 - 1 < 0$)
\item
There must be at least one intersection point of $V_{eff}(\sigma)$
with the $V_{eff}(|\Phi|)=0$ axis.
\end{itemize}

\begin{figure}
 \psfig{figure=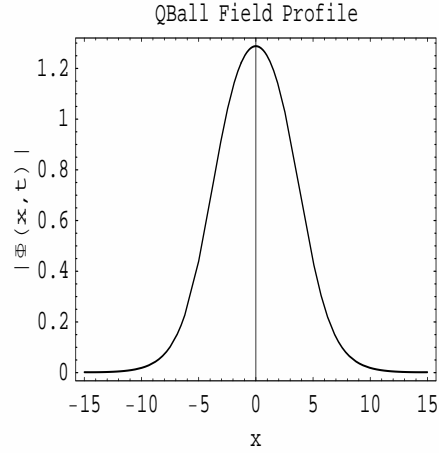,height=3.2in,width=3.2in}
\vspace{-30pt}
\caption{ The field magnitude profile $|\Phi (x,t)|$ of a qball
solution for $B=4/9$, $\omega^2 =0.51$.
   }
\label{fig:phix}
\end{figure}
\vspace{5pt}

The first condition implies $\omega^2 < 1$ while the second
condition implies that the equation $V_{eff}=0$ has at least one
real solution different from 0. It is easy to show that the
constraints imposed on $\omega$ by these two conditions can be
written as

\be
1 - {2\over {9B}} < \omega^2 < 1. \label{constrom} \ee

The energy density of the system (\ref{model}) is
\be
{\cal E}=|{\dot \Phi}|^2 + |\Phi '|^2 + |\Phi|^2 -{2\over 3}
|\Phi|^3 + {B\over 2} |\Phi|^4 \ee
which with the qball ansatz (\ref{ansatz}) becomes

\be
{\cal E}=\sigma'^2 + (1+\omega^2)\sigma^2 - {2\over 3} \sigma^3 +
{B\over 2} \sigma^4. \label{energy} \ee

Our main goal is to study numerically the dynamics and the
stability of single and multi-qball configurations in the context
of the simple model (\ref{model}). The first step in that
direction is to solve equation (\ref{qeq}) numerically (in a
parameter region where qball solutions exist), to find the single
qball profile and then evolve numerically the solution according
to (\ref{feq}) checking total energy and charge conservation.

Fig. \ref{fig:phix} shows the numerically obtained qball profile
for $B=4/9$ and $\omega^2 = 0.51$ using a fourth order Runge-Kutta
scheme. The evolution of this configuration in time when used as
initial condition in equation (\ref{feq}) with periodic boundary
conditions has been performed and found to correspond to a stable
configuration. The total energy and charge Q were conserved to
within about 3\% during the evolution.

Static scalar field configurations in space dimensions higher than
two are unstable towards collapse because both the gradient and
the potential terms of the energy may be shown to decrease with a
rescaling of the configuration corresponding to
collapse\cite{d64}. On the other hand the stability of the
stationary (time dependent) qball configuration towards collapse
or expansion is a result of the different behavior of kinetic
(time dependent) and potential terms towards spatial rescaling  of
the field configuration.  This fact may be seen even in 1+1
dimensions by expressing the energy as a sum of kinetic, gradient
and potential terms \be E=I_1 + I_2 + I_3, \ee where

\ba I_1 &=& \int_{-\infty}^{+\infty}dx \; \omega^2 \; \sigma^2 =
{Q^2 \over {\int dx \; \sigma^2}}, \\
 I_2 &=&
\int_{-\infty}^{+\infty}dx \; \sigma'^2, \\
 I_3 &=&
\int_{-\infty}^{+\infty}dx \; (\sigma^2 -{2\over 3}\sigma^3 + {B
\over 2} \sigma^4). \ea
 After a rescaling of the spatial coordinate $x \rightarrow \alpha
 x$ the expression of the total energy becomes
 \be
 E_\alpha ={1\over \alpha} (I_1 + I_2) + \alpha I_3.
 \ee
For stability we demand that the energy is minimized for
$\alpha = 1$ which implies that
  \be
  {{dE_\alpha} \over {d\alpha}}|_{\alpha = 1} = 0 \Rightarrow  I_1 + I_2 = I_3
  \label{virial} \ee
while the second derivative is ${{d^2 E_\alpha} \over
{d\alpha}^2}|_{\alpha = 1}=2\: (I_1 + I_2) > 0$, implying energy
minimum and therefore stability. Thus the stability of the qball
configuration towards coordinate rescaling  implies the validity
of the virial theorem of equation (\ref{virial}) connecting the
potential energy with the gradient energy and the kinetic energy
(for a similar virial theorem see Ref. \cite{k97}).    We have
confirmed numerically the validity of the virial theorem
(\ref{virial}) and the virial ratio ${{I_1 + I_2 - I_3} \over {I_1
+ I_2 + I_3}}$ was found to be zero to within less than 1\% for
various values of $\omega$ where qball solutions exist and for
various $B$ in the range $2/9 < B < 1$.

In order to study the stability of the qball configuration under
small fluctuations that conserve the charge Q of equation (\ref{charge})
we must consider variations of the energy obtained from the density
(\ref{energy}) and expressed in terms of Q.
  \be
  E= \int dx \; [\sigma'^2 + \sigma^2 - {2\over 3} \sigma^3 +
{B\over 2} \sigma^4] + {Q^2 \over {\int dx \; \sigma^2}}.
\label{enq} \ee
Since the qball is a stationary solution of the
field equations, the first variation of (\ref{enq}) with respect
to $\sigma$ vanishes identically. The second variation may be
written as

\begin{figure}
\hspace{20pt} \psfig{figure=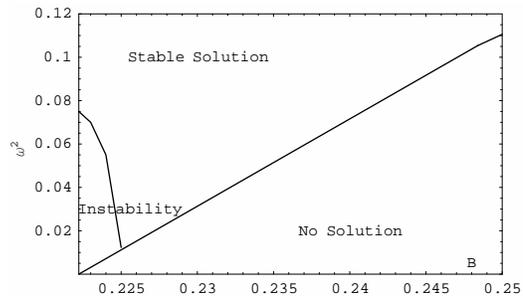,width=3.4in} \vspace{-20pt}
\hspace{-5pt} 
\caption{  Parameter sectors of existence and stability for qball
solutions.
   }
\label{fig:stability}
\end{figure}
\vspace{20pt}

 \be  \delta^2 E = \int dx \; \delta \sigma {\hat O}
\delta \sigma, \ee
where

\be  {\hat O} = -{{d^2} \over {dx^2}} + (1 - 2 \sigma_0 + 3 B
\sigma_0^2) + 3 \omega^2,
\ee
and $\sigma_0$ is the unperturbed
qball solution. For stability we require  $\delta^2 E > 0$ for all
fluctuations that conserve charge. This is equivalent to demanding
that the Hermitian operator ${\hat O}$ has no negative
eigenvalues. In order to find the parameter region corresponding
to stability (no negative eigenvalues) we have solved the
differential equation
\be
{\hat O} \delta \sigma = 0 \label{eig}\ee with initial conditions
$\delta \sigma (0) = 1$ and $\delta \sigma'(0) = 0$ using a fourth
order Runge-Kutta scheme. By varying the parameters $B,\; \omega$
we have identified the line in parameter space where the
eigenvalue problem  (\ref{eig}) has a ground state with zero
eigenvalue. This line clearly separates the parameter sector of
stability from the corresponding sector where
the equation (\ref{eig}) has negative eigenvalues and
the qball is unstable.

Fig. \ref{fig:stability} shows part of the parameter space divided
in three sectors: the sector where stable qball solutions exist,
the sector where qball solutions exist but they are unstable and
the sector where no qball solutions exist because the condition
(\ref{constrom}) is violated. The scale of the plot was chosen in
order to magnify the sector of instability which would otherwise
be too small to be seen.

We have tested and verified the validity of these sectors by
simulating numerically the evolution of an isolated qball using
(\ref{feq}) in the instability and in the stability sectors.

\begin{figure}
\psfig{figure=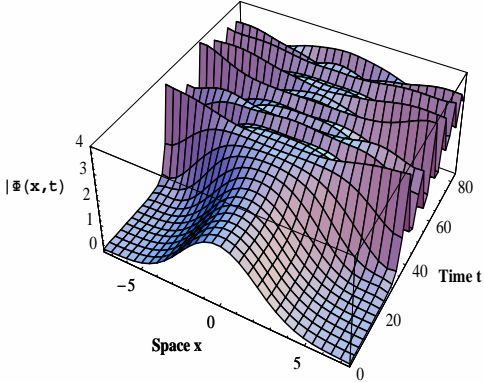,height=3.2in,width=3.2in} \vspace{-35pt}
\caption{Evolution of a qball solution in the instability sector.
The evolution of the field magnitude $\Phi(x,t)$ is plotted. The
time evolution was performed in the                      range
$-7.5<x<7.5$.} \label{fig:evolinst}
\end{figure}

Fig. \ref{fig:evolinst} shows an  example of evolution of a qball 
in the instability sector. On the other hand we have also verified 
stability by performing the time evolution of the field amplitude 
for a qball in the stability sector ($\omega^2 = 0.1 \; B= 2/9 
\simeq 0.222$). The qball was seen to remain stable and evolved
undistorted in time even though the evolution lasted for ten 
internal frequency periods ($t_{max} = 10 ({{2 \pi} \over 
\omega}$)). Fig. \ref{fig:evolinst} shows the same numerical 
experiment for parameters in the instability sector ($\omega^2 = 
0.05 \; B= 2/9 \simeq 0.222$) where equation (\ref{eig}) has 
negative eigenvalues. The evolution lasted only three internal 
frequency periods ($t_{max} = 3 ({{2 \pi} \over \omega}$)) but the 
qball configuration got rapidly distorted and decayed into plane 
waves thus verifying the semi-analytical prediction for 
instability. 

\section{Interactions - Scattering}

We now consider multi-qball configurations in order to study their
interactions and their scattering properties using mainly
numerical simulations of evolution. For a two qball initial
configuration we use the ansatz \be  \Phi (x,t)= \sigma_+ e^{i
\omega_+ t} + \sigma_- e^{i \omega_- t} \label{ansatz2}\ee where
$\sigma_+ \equiv \sigma_1 (x+x_0)$ and $\sigma_- \equiv \sigma_2
(x-x_0)$ with $\sigma_1$ ($\sigma_2$) the field magnitude for a
single qball with $\omega = \omega_+$ ($\omega = \omega_-$).

The interaction potential between the two qballs can be obtained
by subtracting the energy densities $\epsilon_+ (x,t)$,
$\epsilon_- (x,t)$ of each non-interacting qball from the energy
density of the total interacting configuration. After a
straightforward calculation we obtain the interaction energy as

\begin{figure}
\hspace{20pt} \psfig{figure=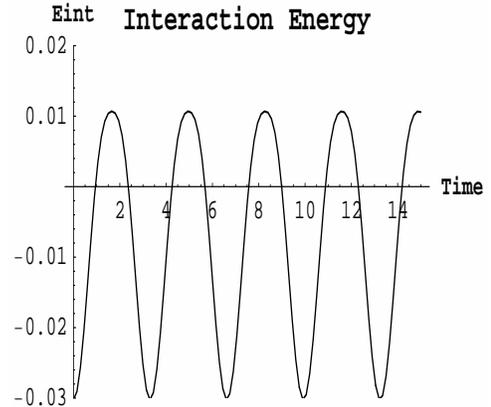,height=3.2in,width=3.2in}
\vspace{-30pt} \caption{The interaction energy of a pair of qballs
calculated using the expression of ${\cal E}_{int}$ with $B=4/9$,
$\omega^2 = 0.9$ showing a behavior of the form $-\cos(2\omega
t)$.
   }
\label{fig:inten}
\end{figure}
\vspace{10pt}

\bea {\cal E}_{int} &=&  \sigma_+ \sigma_- \cos [(\omega_+ -
\omega_-)t] \; F(\sigma_+,  \sigma_-,\omega_+,\omega_-)\\& & +
\sigma^2_+ \sigma^2_- \cos^2 [(\omega_+ - \omega_-)t] +
G(\sigma_+,\sigma_-)
\eea
where $F$ and $G$ are time
independent functions of $O(1)$ and $O(\sigma_+ \sigma_-)$
respectively. It is easy to see from the field equation
(\ref{qeq}) that the fields $\sigma_\pm$ decay exponentially at
infinity and therefore the term proportional to $F$ dominates in
the interaction energy. We therefore expect that the two qball
system will perform oscillations with characteristic angular
frequency $\omega_+ - \omega_-$ while slowly drifting due to the
effect of the small constant interaction term $G$. Indeed this is
what we see in the numerical simulations of the time evolution of
the system.

We have considered the ansatz (\ref{ansatz2}) with $\omega_+ =
\omega_{int}$ and $\omega_- = - \omega_{int}$ which implies
$\sigma_1 (x) = \sigma_2 (x)$ since the internal angular
frequencies of field rotation are opposite. The interaction energy
of this system obtained numerically for $\omega_{int} = 0.9$ as
described above by using the numerically calculated $\sigma (x)$,
is time dependent and is shown in Fig. \ref{fig:inten}.

This interaction energy however assumes that the form of the
initial ansatz is retained during the time evolution of the system
and should therefore be subject to test by numerical simulation.
As expected from the expression of ${\cal E}_{int}$ the time
dependence of the ansatz interaction is proportional to $\cos (2
\omega_{int} t)$.

\begin{figure}
\hspace{20pt} \psfig{figure=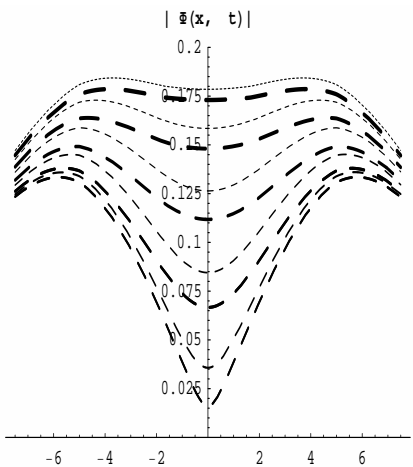,height=3.2in,width=3.2in} 
\vspace{-25pt} \caption{The evolution of the field magnitude 
corresponding to a pair of qballs calculated using the ansatz 
(\ref{ansatz2}) with $B=4/9$, $\omega_{int}^2 = 0.9$.  As time 
increases the line becomes thicker and the dashes larger. The time 
range for a complete period of spatial oscillations is $0<t<3$ 
which implies $\omega_{sp}\simeq 2.1 \simeq 2 \omega_{int}$ to 
within 10\%.} \label{fig:evol2} 
\end{figure}

\begin{figure}
\psfig{figure=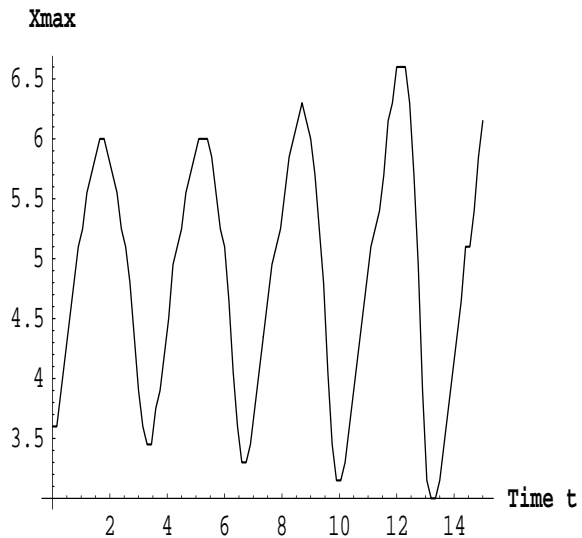,height=3.0in,width=3.0in} \vspace{10pt}
\caption{ Spatial oscillations of the maximum of the field
magnitude due to qball interactions.} \label{fig:spatosc}
\end{figure}

In order to test if the predicted time dependence of the
interaction is realized in a realistic system we have performed a
numerical simulation of the evolution of the ansatz
(\ref{ansatz2}) by solving equation (\ref{feq}) using a leapfrog
algorithm, periodic boundary conditions and initial conditions
based on (\ref{ansatz2}). The parameter values were $B=4/9$,
$\omega^2_{int}=0.9$ and the initial qball distance was $2 x_0
=10$ in a lattice of $2 d = 30$ (these are dimensionless due to
the rescaling of the Lagrangian). The evolution of the field
magnitude $|\Phi (x,t)|$ is shown in Fig. \ref{fig:evol2} where
the oscillations of the system can be seen clearly. The period and
the angular frequency of these oscillations can be found by
plotting the location of the field maxima as a function of time.
This is shown in Fig. \ref{fig:spatosc} where we plot the location
$x_{max} (t)$ of the maximum of  $|\Phi (x,t)|$ for $x>0$ and
$0<t<15$. Clearly the location of the qball at $x>0$ described by
$x_{max}(t)$ performs oscillations due to the interactions with
the qball at $x<0$. The period of these oscillations can easily be
seen from Fig. \ref{fig:spatosc} to be approximately $T \simeq
3.5$ which is consistent with the anticipated result $T = {{2 \pi}
\over {2 \omega_{int}}}\simeq 3.2$ based on the interaction
potential ${\cal E}_{int}$ \ie the spatial frequency of qball
oscillation $\omega_{space}$ is double the internal frequency
$\omega_{int}$ of field rotation.

We have verified this consistency between the analytically
predicted period of spatial oscillation and the one seen in the
simulations for several values of internal rotation frequency
$\omega_{int}$.

\begin{figure}
\psfig{figure=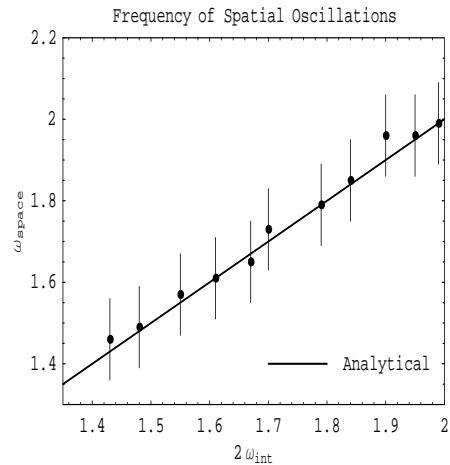,height=3.2in,width=3.2in} \vspace{-30pt}
\caption{The frequency of spatial oscillations of qball position
vs the internal frequency of field rotation.}
\label{fig:spatvsint}
\end{figure}

The result is shown in Fig. \ref{fig:spatvsint} where we plot the
observed angular frequency $\omega_{space}$ of spatial qball
oscillation vs double the corresponding frequency $\omega_{int}$
of internal field rotation. The data points can be fitted well by
a straight line of slope unity as anticipated by the above
described analytical considerations. During all the simulations
the energy and total charge of the system were conserved to within
about 3\%.
The amplitude of the spatial oscillations was not found to be
constant but the system slowly drifted to larger qball separations
with a rate dependent on the parameter values. This behavior is
consistent with the form of the interaction potential ${\cal
E}_{int}$ which includes a subdominant time independent term.

\begin{figure}
\psfig{figure=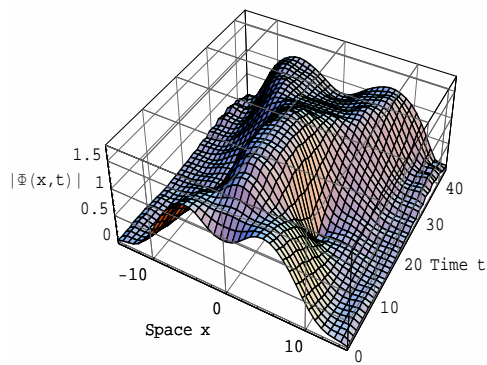,height=3.2in,width=3.2in} \vspace{-40pt}
\caption{Collision of qballs with identical charges and
$v_{01}=-v_{02}=0.2$.} \label{fig:col1d}
\end{figure}

In order to study the scattering of qballs we need to consider
boosted qball configurations \ie qballs moving with a velocity
$v$. Starting from a qball field configuration $\Phi_{v=0} (x,t) =
\sigma(x) e^{i \omega t}$  which does not move in space we can
construct a configuration $\Phi_{v=v_0} (x,t)$ that describes a
qball moving with velocity $v_0$ by performing a Lorenz transform
to the spacetime variables. Let $\gamma = {1\over
{\sqrt{1-v_0^2}}}$ (we use units where the velocity of light c is
unity)  be the Lorenz factor and let
\ba x'&=& \gamma (x-v_0 t)\\
t' &=& \gamma (t- v_0 x)
\ea
be the boosted spacetime variables.
The field configuration $\Phi (x',t')$ expressed in terms of $x,\;
t$ describes a qball moving with velocity $v_0$ \ie $\Phi_{v=v_0}
(x,t)= \Phi (x',t')$.
The initial ansatz for a two qball system prepared for a collision
process may be written as
\be
\Phi (x,t) = \Phi_{v=-v_0} (x-x_0, t)_{\omega_1} + \Phi_{v=v_0}
(x+x_0, t)_{\omega_2}.
\ee
An evolved system of this type with $v_0
= 0.2$ is shown in Fig. \ref{fig:col1d} where we have used
$\omega_1 = \omega_2 = \sqrt{0.51}$ and $B=4/9$. As it can be seen
in Fig. \ref{fig:col1d} the qball collision
results in the
formation of a long lived central qball which subsequently decays
to qballs similar to the original ones.

\vspace{-20pt}

A more interesting evolution occurs in two space dimensional
systems which will be described in the next section. There it will
be seen that the effect of right angle soliton scattering which
has been observed in topological solitons persists also in the
case of the nontopological qballs in a generalized form.

\section{Dynamics in Two Space Dimensions}
The model (\ref{model}) is extended to 2+1 dimensions by
letting the index $\mu$ take the values $\mu=0,1,2$. The details
of the formalism are similar to those given in Sec. II. That is,
the rescaled form of the Lagrangian is given by
Eq.~(\ref{modelresc}) while the field equation reads
\be
{\ddot \Phi} - \Delta \Phi + \Phi - |\Phi| \Phi + B |\Phi|^2 \Phi
= 0, \label{feq2D} \ee where $\Delta = \D^2_x+\D^2_y$.
We look for
axially symmetric solutions of Eq.~(\ref{feq2D}) and make the
qball ansatz
\be
\Phi (\rho,t) = \sigma (\rho) e^{\pm i \omega t},
\label{ansatz2D}
\ee
where $\rho = \sqrt{x^2+y^2}$. The profile $\sigma(\rho)$ satisfies
\be
\sigma'' + {\sigma' \over \rho} + (\omega^2 -1) \sigma + \sigma^2
   - B \sigma^3 = 0.
\label{qeq2D}
\ee

\begin{figure}
   \psfig{file=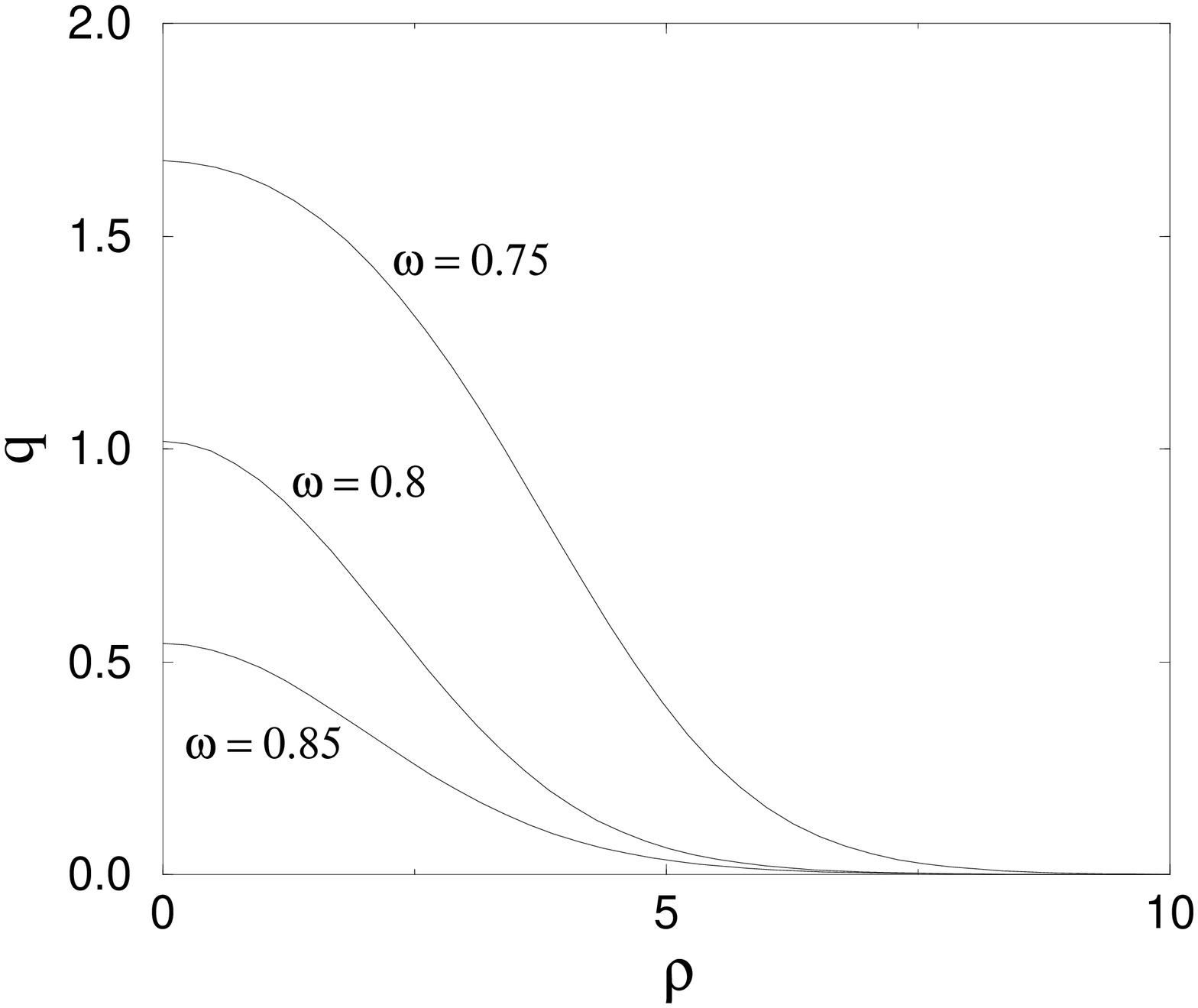,width=7.0cm,angle=-90}
   \vspace{15pt}
   \caption{The numerically calculated profile of the qball solutions
         for $B=4/9$ and various frequency values.
          In the vertical axis we plot the charge density
       $q = \omega\,\sigma^2$.
Qballs exist for $0.5 < \omega^2 < 1$ or $0.71 \lsim \omega <1$.}
   \label{fig:profile}
\end{figure}

\noindent The boundary conditions which have to be met are
given in Eqs.~(11) and (12). Eq.~(\ref{qeq2D}) is expected to
have localized solutions by arguments which are presented in
\cite{c85} and are similar to those used for the 1D model
of Sec. II.

The search for the solutions is here not straightforward as for
the 1D model (Figs. 1,2). In the present two dimensional case,
solutions are found by a
numerical shooting method. Fig.~\ref{fig:profile} shows the
profile of the calculated qballs for $B=4/9$ and for various
values of the frequency $\omega$. We represent the qball profile
through the charge density $q= \omega\, \sigma^2$ (cf.
Eq.~(\ref{charge})).

The virial theorem mentioned in Sec.~II has, for our two dimensional
theory, the form
\be
 I_3 = I_1,
\ee where the symbols are defined as the two dimensional analogues
of (18) and (20). The virial relation is used to check the precision of
our numerical calculations. Indeed, the solutions that we find by
the shooting method satisfy the above virial relation to very good
precision, better than 1\%.

Finally, we note that our shooting
method is unable to find the qball profile for the whole frequency
range (\ref{constrom}).
 Numerical errors render the method
inapplicable near the lower $\omega$-bound. For instance, for
$B=2/9$ we are able to find the solutions with $0.035 < \omega^2 <
1$ and for $B=4/9$ we find those in the frequency range $0.52 <
\omega^2 < 1$. The difficulties with the numerics should be
anticipated. Their origin can be traced to the theoretical
arguments of \cite{c85} with respect to the existence of qballs in
dimensions higher than one. It is actually the friction term in
Eq.~(\ref{qeq2D}) which is responsible for the numerical difficulties.

We further note that a steadily moving qball can
be found by applying a Lorenz boost to a static one as discussed
in the previous section. The calculated qballs will be used in the
following in numerical simulations on a two-dimensional lattice.
We shall first study their stability. Then we shall study
interactions between two qballs, in particular we shall perform
simulations of scattering.

The numerical mesh we use is typically $300\times 300$ and the
lattice spacing is 0.3. Such a numerical mesh is appropriate to
accommodate the structures given in Fig.~\ref{fig:profile} and
provide good resolution. The time evolution is performed by a
fourth order Runge-Kutta method.

First, we put a single qball at the center of our numerical mesh
and simulate its time evolution in order to test its stability. We
use $2/9 < B < 1$ and test all the values of $\omega$ for which we
are able to find the qball profile by our shooting method. We find
that qballs are stable everywhere in the $B-\omega$ plane and they
travel undistorted with the given constant velocity.

In the next set of simulations we discuss the problem of
interaction between qballs. Specifically, we focus  on
simulations of scattering between two qballs.
Such simulations have been performed and have proved to be fruitful for
topological solitons. We shall perform all our subsequent
numerical simulations using the values $B=4/9$, $\omega = \pm 0.75$.

We consider a 2D generalization of the
ansatz (30) which represents two qballs set in a
head-on collision course. We use $\omega_1=\omega_2=0.75$ and
a small velocity, namely $v_0=0.2$. Qballs are
represented through their charge density:
\be
q = {1 \over 2 i} (\Phi^*\, \D_t \Phi - \Phi\, \D_t \Phi^*).
\label{qdensity} \ee

\begin{figure}
\hspace{15pt} \psfig{file=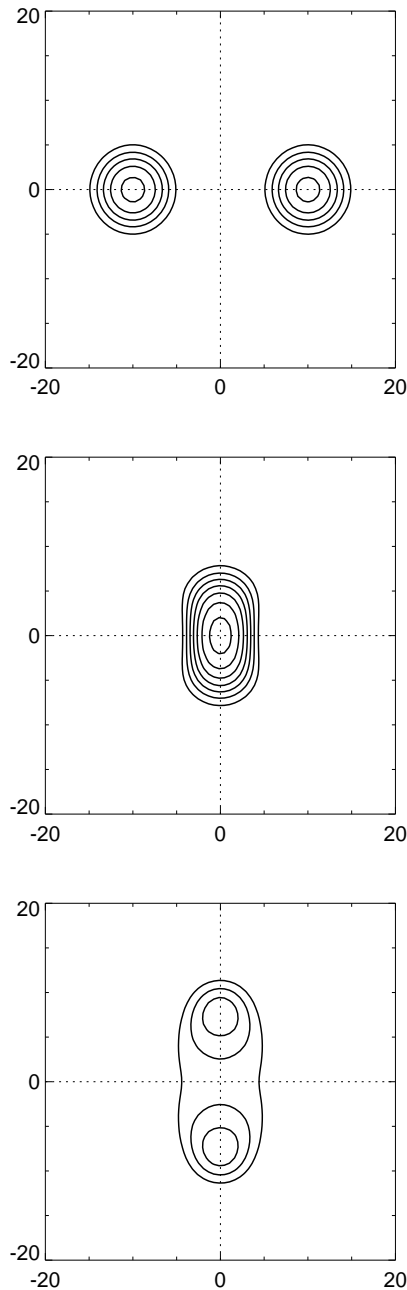,height=6.7in}
\vspace{15pt}
\caption{Head-on collision of two qballs with the
same charge. Contour plots of the charge density are given.
Contour levels: 0.4, 0.7, 1, 1.3, 1.6, 1.9, 2.2. Three
snapshots are shown: initial ansatz (upper
entry, time $t=0$), at collision time (middle entry, $t=41.9$) and
well after collision (lower entry, $t=55.9$). Parameter values:
$B=4/9$, $\omega = 0.75$, initial velocity of each qball
$v_0=0.2$.}
\label{fig:rightangle}
\end{figure}

In Fig.~\ref{fig:rightangle} contour plots are given
for the charge density at three characteristic snapshots of this
numerical simulation. The first entry of the figure gives
the initial ansatz.
The two qballs are initially at a distance of 20 units apart so
that there is no overlap and interaction between them. Subsequently
they collide at the origin (middle entry) and scatter at right
angles (lower entry).

The scattering scenario produced by the computer simulation is
quite interesting. Indeed, this dynamical behavior has been found
to be a robust feature in a variety of models in two space
dimensions which have topological soliton solutions
\cite{ruback,ward95}. However, in our case there is no topological
invariant associated to the qballs, so we need to discuss it
further.

The crucial observation is that
the underlying dynamics which induces this type of scattering can be
attributed solely to the Hamiltonian structure of the model. It is
related to the conservation laws of our model and in particular to
the linear momentum conservation \cite{komineas}. We have followed
the linear momentum density plot along the lines described in
\cite{komineas} and we have found a behavior similar to that of topological
solitons at the collision time. As a conclusion,
despite that in our case there is no topological invariant
associated to the qballs, the dynamics
related to the right angle scattering behavior is
the same as that for their topological counterparts.
We should therefore expect such a dynamical behavior to
occur generically. We shall study in this section the
validity of the above remarks.

\begin{figure}
   \hspace{15pt} \psfig{file=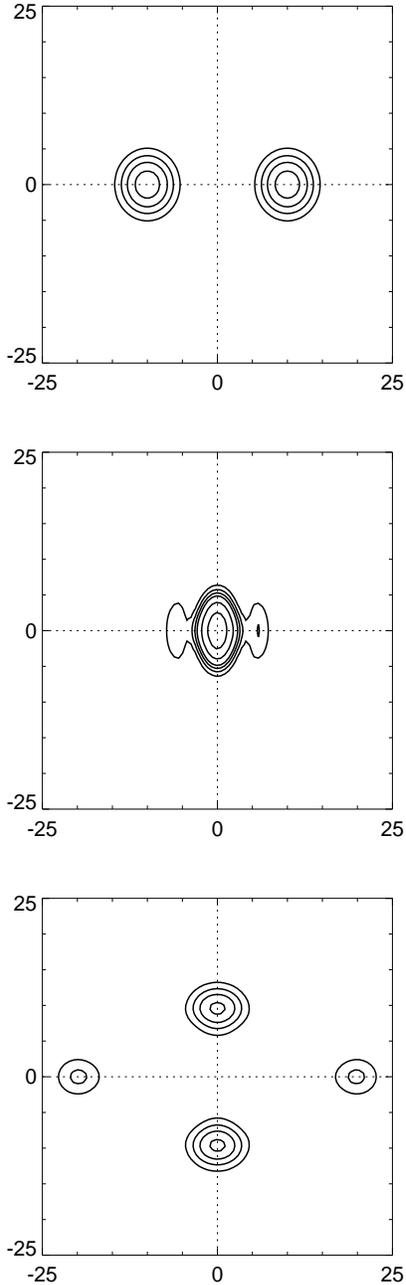,height=6.7in}
   \vspace{15pt}
   \caption{Head-on collision of two qballs with the same charge.
Contour plots of the charge density are given.
Contour levels: 0.4, 0.8, 1.2, 1.6, 2.4, 3.2.
Snapshots at: $t=0$ (upper entry),
$t=19.6$ (middle entry)  and $t=69.8$ (lower entry).
Initial velocity of each qball $v_0=0.4$.
Rest of parameters as in Fig.~\ref{fig:rightangle}.
   }
   \label{fig:fourqballs}
\end{figure}

Following the simulation for later times, the two qballs are seen
to eventually stop along the $y$-axis. They then attract, collide
again and subsequently scatter and re-emerge on the $x$-axis. This
scenario goes on and gives an oscillating system with
successive right angle scattering. We shall not pursue further
here this oscillating behavior.

The next simulation we present has been prepared in a way
similar to the previous one, but now the initial 

\begin{figure}
   \hspace{15pt} \psfig{file=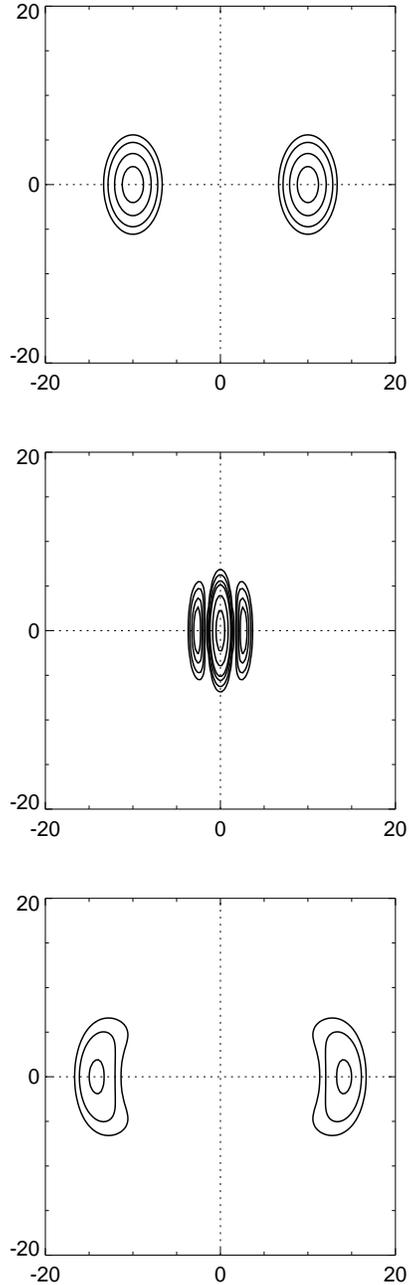,height=6.7in}
   \vspace{10pt}
   \caption{Head-on collision of two qballs with the same charge.
Contour plots of the charge density are given. Contour levels:
0.4, 0.8, 1.6, 2.4, 5, 8. Snapshots
at: $t=0$ (upper entry), $t=12.6$ (middle entry)  and
$t=29.3$ (lower entry). Initial velocity of each qball $v_0=0.8$.
Rest of parameters as in Fig.~\ref{fig:rightangle}
   }
   \label{fig:forward}
\end{figure}
\vspace{-15pt}

\noindent velocity of the
qballs is set to a higher value $v_0 = 0.4$. The results are shown
in Fig.~\ref{fig:fourqballs}. At the initial stages, the
simulation does not differ from our first simulation. At collision
time (middle entry) a central qball is formed.
However, the subsequent evolution is considerably
different. In the lower entry of the figure we have two qballs
which have evolved from a right angle scattering process. They are
located on the $y$-axis and are almost static. In addition to
that, two other qballs are continuing their route along the
$x$-axis and they are drifting away from each other. Their drift
velocity is approximately 0.24.

It would be interesting to have some insight in the above
described process. The basic remark is that the
symmetry that has lead to the right angle
scattering in the first (slow qballs) case, leads here again to
the same result. On the other hand, the high kinetic energy of the
initial qballs allows for more complicated scenarios like the
formation of new qballs. That is, some amount of energy that does
not follow the right angle scenario, reorganizes to form qballs on
the $x$-axis. In the case of topological soliton interactions,
arguments related to topology would usually preclude such
scenarios. These arguments do not apply in the case of qballs
thus we are led to a novel result. An interesting observation is that the
four solitons which are created after the collision seem to be
qballs with a frequency different than that of the initial ones.
This can be seen by comparing the last entry of
Fig.~\ref{fig:fourqballs} with the profiles of qballs of various
frequencies, obtained by our numerical shooting method.

We have repeated the above scattering numerical experiment with qballs of
higher initial velocity $v_0=0.8$. The results of the simulation
are given in Fig.~\ref{fig:forward}. A new type of
collision different than in the two previous cases arises. The qballs
actually collide to form a central qball at the origin
(middle entry of the figure). However,
after the collision (lower entry of the figure) two qballs
re-emerge traveling along the $x$-axis and drifting away from each
other. Their drift velocity after collision is approximately 0.75.
No right angle scattering is present in this case so the qballs
appear to be almost non-interacting. This result is dramatically
different from what we see in the case of the slow qballs of
Fig.~\ref{fig:rightangle} where pure right angle scattering
occurs. It resembles however similar results discussed in the
literature for scattering of topological solitons moving with very
high relativistic velocities\cite{shellard}.

Notice also that Ward's chiral model, which is integrable in 2+1
dimensions, presents in some respects similar dynamical behavior.
Specifically, it has solutions which correspond to the right angle
scattering \cite{ward95} and others which represent noninteracting
solitons \cite{ward88} although the velocity of the solitons does
not seem to play any role there. On the other hand, the type of
scattering shown in Fig.~\ref{fig:fourqballs} seems to be specific
to nontopological solitons and, to our knowledge, it has not been
observed in other two-dimensional models.

The right angle scattering of solitons has also been observed in a
three-dimensional Yang-Mills-Higgs theory \cite{manton}. The
problem has been approached within the moduli space, that is, the
space of the multi-monopole solutions in the Bogomonly limit. It
has been stressed that the scattering behavior of the monopoles
can be understood when they are considered almost static during
the interaction. This requires that the velocity of the
interacting monopoles is small. Such is certainly the case with
the first of our simulations here shown in Fig~\ref{fig:rightangle}.

\begin{figure}
   \hspace{15pt} \psfig{file=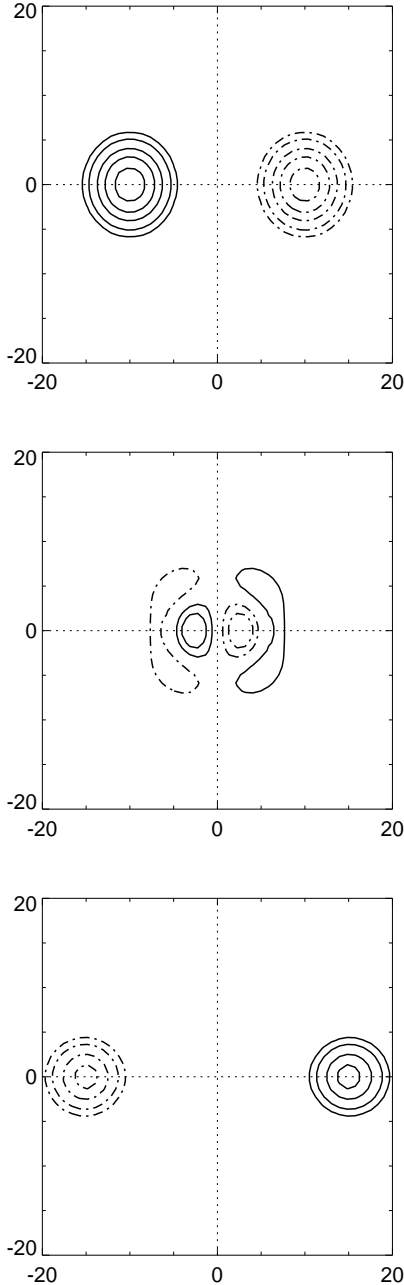,height=6.7in}
 \vspace{25pt}
  \caption{Head-on collision of two qballs with opposite charge.
Contour plots of the charge density are given. Solid
lines represent positive values and dashed-dotted lines negative
values. Contour levels: $\pm 0.2,\pm 0.4,\pm
0.8,\pm 1.2,\pm 1.6$. Snapshots at:
$t=0$ (upper entry),
$t=28$ (middle entry)  and $t=61.4$ (lower entry).
Initial velocity of each qball $v_0=0.4$. Rest of parameters as in
Fig.~\ref{fig:rightangle}
   }
   \label{fig:opposite}
\end{figure}

Additional insight in the dynamics can be gained by looking at
collisions between qballs of opposite charge. Such are two qballs
with the same profile and opposite field precession frequency (cf.
Eq.~(\ref{ansatz2D})). Notice that the arguments of \cite{komineas}
can not be applied in this case, at least not in a straightforward
manner. In particular, these arguments indicate that we may not
expect that the two {\it initial} qballs can re-emerge after
collision, traveling at a direction perpendicular to the initial
one. Rather, we would have to think of a combination of parts of
the initial qballs which would form the final solution after
collision. Such a combination, involving parts of qballs with
opposite charge, does not seem to form any solution
of the present theory, therefore we don't expect a right angle scattering
behavior in the current case. We have to resort to numerical
simulations in order to determine the actual behavior of the system.

We now use an initial ansatz appropriately chosen to account for
opposite-charge qball collision. This ansatz is shown in the upper
entry of Fig.~\ref{fig:opposite} through contour plots of the
charge density. The positive-charge qball is located on the left
of the figure and the negative-charge one on the right. We apply
to each of the initial qballs a Lorenz boost with a moderate
velocity $v_0=0.4$. The middle entry of the figure shows that,
quite interestingly, at the time of collision a mixed state is formed
between the two qballs. After that, the two qballs separate again,
they appear to pass through each other and drift apart. The final
qballs are not identical to the initial ones. The charge Q of each
of them is approximately half of that of the initial state and it
corresponds to a frequency $\omega \sim \pm0.6$. They have also
decelerated and their mean velocity is approximately equal to
0.25.

In Sec.~II, we have found an interaction between the pair of oppositely
charged qballs which introduces an oscillation  frequency to the
system equal to twice the qball internal frequency.
This  interaction is present here and
it manifests itself in oscillations of
the magnitude of the field values and also in oscillations of the
position of the qball centers. The oscillations appear when the
qballs are coming close to collide. They also persist after the
collision even when the qballs are well separated (\ie at the
instant of the last entry of Fig.~\ref{fig:opposite}) and they don't
show any tendency to fade out.

Computer plots of the distribution of the energy density of the
system shows that some small amount of energy is dissipated at
right angles after collision. The underlying mechanism for this
phenomenon is presumably similar to that leading to right angle
scattering in the first of our simulations in this series. Similar
behavior has been observed for two colliding
topological solitons with opposite topological charge \cite{shellard}.
In this case, the solitons annihilate and the energy is
dissipated at right angles.

The scattering behavior described by the simulations of this section can be
viewed as a generalization of the corresponding behavior of
topological solitons. We observe the two extreme cases
which have also been observed in the topological case
(pure right angle scattering at low
velocities and pure forward scattering at very high velocities)
but we also observe an intermediate case of a qball split to a
forward and a right angle scattering component at intermediate
velocities. The forward scattering component gets amplified at
high collision velocities. This amplification occurs at the
expense of the right angle scattering component.

\begin{figure}
\psfig{figure=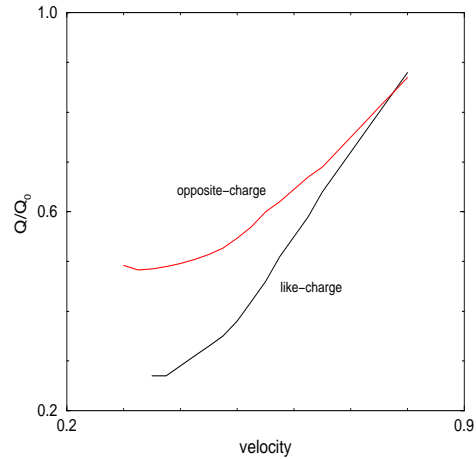,height=2.8in,width=2.8in,angle=-90}
\vspace{10pt}
\caption{The fractional charge in forward scattering of qballs as
a function of the rest frame qball velocity. The forward component
is clearly amplified at large velocities. Points in the figure are
plotted every 0.25 velocity units. Then the points are connected
by straight lines.}
\label{fig:qq0}
\end{figure}
\vspace{15pt}

The rich behavior of qballs in the scattering simulations calls
for an overview of the different phases.
\begin{itemize}

\item
Low velocity scattering of qballs with the same charge leads to
initial right angle scattering and a bound system that performs
breather type oscillations.
\item
Intermediate velocity scattering ($0.3<v<0.7$) leads to a
combination of forward and right angle scattering and the forward
scattered qballs drift away from each other and escape to
infinity.
\item
High velocity scattering ($v>0.7$) leads to pure forward
scattering. The forward scattered qballs drift away from each
other and escape to infinity.
\end{itemize}

To clarify the transition from pure right angle scattering to pure
forward scattering in the case of identically charged
non-topological solitons we plot (Fig.~\ref{fig:qq0}) the charge
of each of the forward scattered qballs as a fraction of the
initial qball charge, versus the initial velocity
\footnote{ In
order to have clearly separated qballs in the charge computation,
Fig.~\ref{fig:qq0} shows results only for $v>0.3$ and $v>0.35$ for
the opposite and same-charge case respectively. Then the two
qballs clearly separate from each other after collision and drift
away. For smaller velocities, the qballs collide, go through each
other, but do not subsequently drift away. Instead the system
remains bounded and performs breather-type oscillations.}. 
Since
the total charge is conserved the rest of the charge is scattered
at right angles. A small amount is dissipated throughout our
lattice.

The charge of the outgoing qballs is computed as follows: We let
the qballs travel 30 space units on the $x$-axis, after the
collision and we integrate the charge density within a disc  of
radius 15 space units around each qball center. This result is
considered to be the charge of the final qballs.

In the case of scattering of oppositely charged qballs we have
also observed an increase of the forward scattered component
charge for high velocities (upper curve of Fig.~\ref{fig:qq0}).
This increase however does not occur due to reduced right angle
scattering (this does not happen in this case) but due to reduced
annihilation between the opposite qball charges which is a result
of the reduced time of overlap between the fast moving qball
profiles.

\section{Conclusion - Outlook}

We have studied the stability and the dynamics of qballs in the
context of a simple toy model. We have found that the qball
instability sector is a very small sector in parameter space. The
validity of a virial theorem in the stability sector has also been
verified and we have shown that systems of qball pairs tend to
perform breather-type oscillations for long time periods compared
to the period of the internal field rotation. Finally we have
studied numerically the scattering of qball pairs and found that
qballs in two space dimensions with the same charge tend to
scatter at right angles and subsequently perform oscillations with
repeated collisions through right angle scattering. At
relativistic collision velocities a significant forward scattering
component has also been found which gets amplified as the
collision velocity increases. This effect is consistent with
numerical experiments involving topological solitons where the
charge is discretized due to topology. In that case complete
forward scattering is found for $v>v_{crit}\simeq
0.9$\cite{shellard}.
In the nontopological case the smooth
interpolation between the above two regimes is allowed because the
charge is not discretized.

These results are interesting not only in the context of the
general study of the solitonic dynamical properties but also in
the context of realistic physical systems. For example in a
cosmological setup where dark matter comes from susy qballs,
breather-type large qball systems could lead to detectable
signatures in the gravitational wave spectrum either by direct
emission or by exciting vibration modes of the neutron stars where
they can be trapped\cite{m98}. The detectable characteristics of
such breather type cosmological systems are currently under
investigation. Another interesting implication is related to the
statistical mechanics of qball systems\cite{k98}. Our results
imply that the number of qballs is not conserved in such systems
and in fact it may increase rapidly during multiple high velocity
collision processes. This is in contrast with the assumptions made
in recent work \cite{ks98} which assumes that merging is the only
possible outcome of qball collisions.

A natural extension of our work is the numerical study of the
formation of qballs during a (cosmological) phase transition. Such
studies based on numerical simulation s\cite{yb90,z96} of a system
though a temperature quench have been performed extensively in the
context of topological solitons \cite{z96,lz97} but not in the
context of nontopological configurations like the ones considered
in the present study.
Mechanisms of thermal creation and annihilation of nontopological
solitons in the aftermath of a phase transition have been
previously studied. For small Q balls it is assummed their fusion
into larger ones to be dominant for both high and low thermal
reaction rates \cite{k88} Clearly our results modify the above
picture in an interesting way.
The numerical simulations performed
here can be extended in a straightforward way to apply to the case
of qball formation during a quench. The realization of this
extension is currently in progress.

\section{Acknowledgements}
We thank N. Papanicolaou for useful conversations

\vfill

\eject

\widetext

\end{document}